# Discovering Neighbor Devices in Computer Network

Development of CDP and LLDP Simulation Modules for OMNeT++

Vladimír Veselý, Tomáš Rajca
Department of Information Systems, Faculty of Information Technology
Brno University of Technology
Brno, Czech Republic
veselyv@fit.vutbr.cz, xrajca00@stud.fit.vutbr.cz

*Abstract*—The purpose of data-link layer discovery protocols is to provide the network administrator with the current information (i.e., various Layer 2 and 3 parameters) about neighbor devices. These protocols are invaluable for network monitoring, maintenance, and troubleshooting. However, they start to play an important role in the operation of data-centers and other high-availability networks. This paper outlines design, implementation and deployment of Cisco Discovery Protocol and Link Layer Discovery Protocol simulation modules in OMNeT++ simulator.

## I. Introduction

Layer 2 discovery protocols have been developed to share information between directly connected devices. They send specific device's information (e.g., device role, interface state, assigned IP address, operating system version, Power over Ethernet capability, duplexness, VLAN configuration, etc.) to neighbors. These protocols are useful during network maintenance and process of troubleshooting when the administrator is trying to locate the source of a problem and isolate its layer presence. Cisco Discovery Protocol (CDP) was the first one from this family of data-link layer discovery protocols. CDP usage is limited to Cisco devices only due to its proprietary nature. Other vendors decided to follow the idea and developed their variants such as Foundry Discovery Protocol by Brocade, Bay Network Management Protocol and Nortel Discovery Protocol by Nortel, Extreme Discovery Protocol by Extreme Networks, Link Layer Topology Discovery by Microsoft, and others. In order to offer a multi-vendor environment, IEEE came with unifying protocol offering the same functionality as above mentioned representatives. It is codified in IEEE standard 802.1AB and called Link Layer Discovery Protocol (LLDP).

One of OMNeT++'s most popular frameworks is INET [1] that aims at providing models for Internet devices, protocols, and a mechanism to help with network design and configuration testing and evaluation. The Automated Network Simulation and Analysis for Internet Environment (ANSAINET) project is dedicated to the development of a variety simulation models compatible with RFC specifications or referential implementations, which extends the standard INET framework.

Both CDP and LLDP are de facto industry standards when it comes to network operation life-cycle. Since our goal is to develop simulation models for various networking technologies, we have decided to extend the functionality of ANSAINET. Hence, the paper outlines processes of adding support for CDP and LLDP, and it should be treated as finalized software contribution rather than research effort.

This paper has following structure. Section II covers a quick overview of existing implementations. Section III describes the operational theory and implementation design notes. Section IV contains testing scenarios. The paper is concluded in Section V, which also outlines our future work.

## II. State of the Art

This section briefly overviews existing CDP and LLDP implementations for hardware/software routers and also simulators.

Since CDP is Cisco's intellectual property, CDP deployment in hardware is limited to Cisco's product portfolio only. Scarce CDP availability exists for simulators too. Cisco Packet Tracer [2] allows CDP configuration since its earliest versions. However, Cisco Packet Tracer is closed and proprietary simulator used mainly as an education tool.

On the other hand, LLDP is supported by a wide range of networking equipment vendors (e.g., Juniper, Hewlett-Packard, Arista, Brocade, including Cisco, and others) and operating systems (both Windows and Unix-based). We are not aware of any CDP/LLDP support by NS2/3 or OPNET.

During the ANSA project run, we have extended available simple network node with additional functionality – support for various routing, switching and data-link layer discovery protocols. The resulting `ANSARouter`, `ANSASwitch`, and `ANSAHost` components are a compound modules integrating all expected functionality in programmable simulation modules that adopt a Cisco-style representation of configuration, textual outputs (e.g., routing table format) and debugging information. This paper discusses CDP and LLDP implementation and their integration as new `networkLayer` submodules to `ANSARouter` and `ANSASwitch`. The simplified schema showing this integration in `ANSARouter` is depicted in Figure 2.

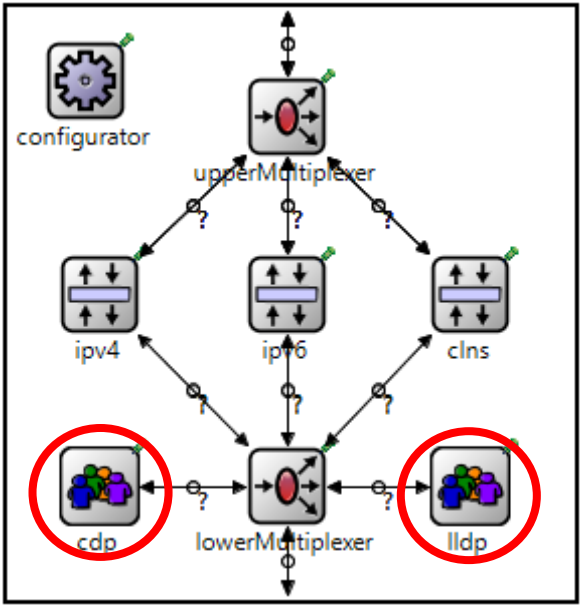

Figure 2: Structure of ANSARouter networkLayer

## III. PRINCIPLES

This section provides a description of principles of both CDP and LLDP. It includes the format of protocol messages and designed abstract data structures.

The reader is advised to follow references in order to learn more about particular protocol. CDP theory is based on references [3], [4], and [5]. LLDP theory is covered in sources [6], [7], [8], and [9].

### A. Cisco Discovery Protocol

The current version 2 of CDP operates on any data-link layer technology with Subnetwork Access Protocol (SNAP) support, i.e., Ethernet, WiFi, Frame Relay, ATM or PPP.

CDP messages are sent to multicast MAC 01:00:0c:cc:cc:cc by default every 60 seconds. Data contained in CDP message are device dependent. CDP message consists of a generic header and a variable number of type-length-value (TLV) triplet fields. CDP header has following three mandatory TLVs – *Version*, *Time to Live*, *Checksum*. Shortened list of TLVs recognized by CDP is summarized in Table V, where columns marked "CDP TLV" and "TLV's Description" are relevant.

In OMNeT++, CDP is implemented as the compound module `CDP` interconnected with `lowerMultiplexer` of `networkLayer`. It consists of four submodules that are depicted in Figure 1 and briefly described in Table I. Our implementation is in full compliance with the observed behavior of Cisco's referential behavior.

### B. Link Layer Discovery Protocol

LLDP operates in logical link control sublayer of data-link layer employing SNAP. LLDP terminology introduces agent, which is LLDP instance bound to a certain device's port. The agent sends and processes LLDP messages on a given interface.

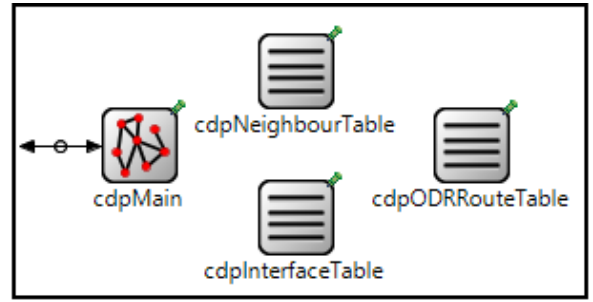

Figure 1: CDP module structure

TABLE I. DESCRIPTION OF CDP SUBMODULES

| Name | Description |
| --- | --- |
| `cdp Main` | This module has core CDP functionality, which is responsible for parsing of XML configuration, message and timers handling, on-demand routing (ODR) functionality. Lightweight ODR is one of the main reasons we decided to implement CDP. |
| `cdp Neighbor Table` | This abstract data structure stores received CDP information from directly connected neighbors. Records are dynamically updated with every new CDP message received and expire after a given Time To Live value. |
| `cdp Interface Table` | Interface table contains a list of CDP enabled interfaces. This table state influences a periodic generation of CDP messages and included data. |
| `cdp ODRRoute Table` | This table holds routes learned via ODR extension. Each route is accompanied just like RIP with Invalid, Holddown and FlushedAfter timers. |

LLDP data are stored in two management information bases (MIB) – first one local (for the device itself), second one remote (for information from neighbors).

LLDP message consists of a header with mandatory TLVs – *Chasis Id*, *Port Id*, *Time To Live* – followed by optional TLVs with additional data. All LLDP TLVs are included in Table V, where columns "LLDP TLV" and "TLV's Description" are relevant to LLDP. Additional TLV sets extending LLDP exist (e.g., LLDP-MED, DCBXP) but they are out of the scope of this paper. Comparing to CDP, LLDP optionally offers error management for its communication. Moreover, LLDP has also built-in rate-limiter for sending based on credit. LLDP standard assigns three dedicated multicast destination MAC addresses 01:80:c2:00:00:00, 01:80:c2:00:00:03, and 01:80:c2:00:00:0e (this one is default for Ethernet-based networks). LLDP message is periodically generated (by default) every 30 seconds.

We have similarly designed LLDP as CDP. `LLDP` compound module implements `INetworkLayerLower` interface, and it is interconnected with `lowerMultiplexer`. The module structure is depicted in Figure 3 and submodules description listed in Table II.

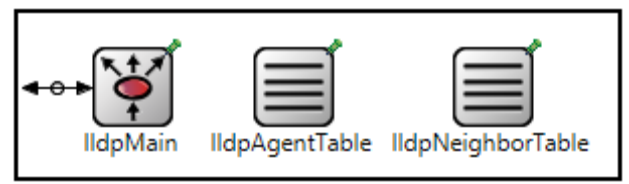

Figure 3: LLDP module structure

TABLE II. DESCRIPTION OF LLDP SUBMODULES

| Name | Description |
|---|---|
| lldp Main | This module delivers core LLDP functionality. It sets up LLDP module based on XML preconfiguration. It governs sending and receiving of LLDP messages. It maintains neighborship and relevant information. |
| lldp AgentTable | Functionality is comparable with cdpInterfaceTable in sense that it contains interface specific LLDP settings. |
| lldp NeighborTable | The functionality of this abstract data structure is analogous to cdpNeighborTable. |

IV. VALIDATION AND VERIFICATION

This section contains information about verification and validation of implemented simulation modules over the same set of scenarios. Demonstration example is purposely too basic, but both protocols have also been verified on more complex topologies.

Verification was conducted using a traditional approach employing code review, debugging and documentation [10]. We have found out that simulation models comply with their corresponding specifications; namely, the format of messages, configuration parameters meaning, and the functionality in all tested cases. In simulation validation, we have measured the accuracy of simulation models to real implementations on Cisco devices. As a part of this activity, we have set up same network scenarios in both simulator and the real environment. As a source of information, we analyzed packets exchanged between devices and debugging outputs of related processes. We built the test-bed environment from Cisco routers running IOS version 15.4(2)T4, Cisco switches running IOS version 15.2, and host stations with Windows 7.

Figure 4 shows the basic topology used for validation. It consists of three ANSARouter instances (marked R1, R2, and R3) and one ANSASwitch instance (marked S1) providing CDP/LLDP functionality and two ANSAHost instances (Host1 and Host2). To compare CDP and LLDP with each other, we have changed default LLDP timers – periodic generation of messages to 60 seconds and *Time To Live* value to 180 seconds.

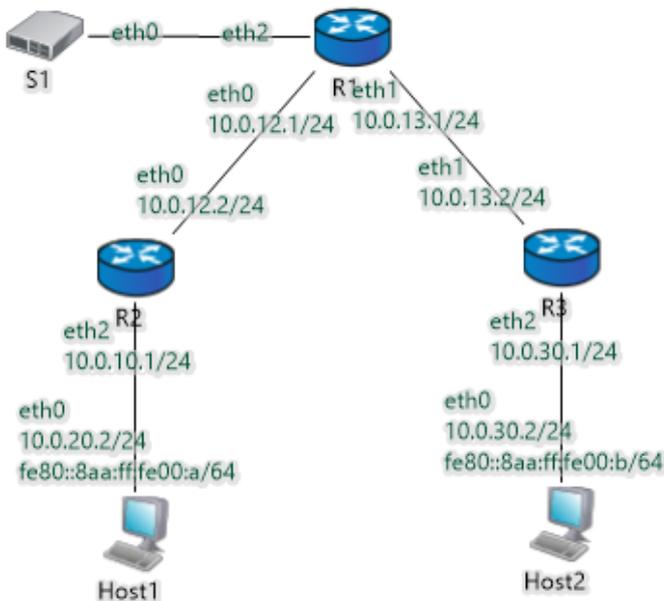

Figure 4: CDP/LLDP testing topology

Both protocol offer fast-start feature, which speeds up the process of neighbor discovery. During the fast-start, periodic message generation interval is just 1 second. Fast-start lasts for: a) three consecutive message updates in case of CDP; b) one to eight (by default three) consecutive message updates in case of LLDP. Fast-starts happens each time when: a) interface restarts in case of CDP; b) MIB content changes in case of LLDP standard; c) a new end-host is detected, or LLDP-MED TLV is exchanged in case of LLDP implementation by Cisco.

*A. Initial Discovery*

This test shows initial neighbor discovery when the interface changes from down state to up state after device successfully starts. We recorded all CDP/LLDP messages into PCAP file and compared timestamps. Table III shows the result for a link between R1 and R2 for both protocols (i.e., CDP and LLDP) and both simulated and real scenario.

TABLE III. TIMESTAMP COMPARISON FOR INITIAL DISCOVERY

| Direction | CDP | | LLDP | |
|---|---|---|---|---|
| | Simul. [s] | Real [s] | Simul. [s] | Real [s] |
| R1 → R2 | 0.000 | 0.300 | 0.000 | 1.600 |
| R2 → R1 | 0.000 | 5.370 | 0.000 | 1.900 |
| R1 → R2 | 1.000 | 1.300 | 1.000 | missing |
| R2 → R1 | 1.000 | 6.370 | 1.000 | missing |
| R1 → R2 | 2.000 | 2.310 | 2.000 | missing |
| R2 → R1 | 2.000 | 7.380 | 2.000 | missing |
| R1 → R2 | 62.000 | 57.550 | 62.000 | 61.300 |
| R2 → R1 | 62.000 | 66.850 | 62.000 | 61.400 |

*B. Interface Restart*

This test tracks events bound to the flapping of interface between R1 and R2. After the link goes down at $t = 50s$, records expire from tables at $t = 180s$. Then at $t = 200s$ connection is reestablished and CDP/LLDP messages are first to appear on the wire. Table IV shows simulated and real scenario result for link between R1 and R2 for both CDP and LLDP.

TABLE IV. TIMESTAMP COMPARISON FOR INTERFACE RESTART

| Direction | CDP | | LLDP | |
|---|---|---|---|---|
| | Simul. [s] | Real [s] | Simul. [s] | Real [s] |
| R1 → R2 | 200.000 | 199.480 | 200.000 | 202.000 |
| R2 → R1 | 200.000 | 201.500 | 200.000 | 205.000 |
| R1 → R2 | 201.000 | 200.500 | 201.000 | missing |
| R2 → R1 | 201.000 | 202.510 | 201.000 | missing |
| R1 → R2 | 202.000 | 201.510 | 202.000 | missing |
| R2 → R1 | 202.000 | 203.510 | 202.000 | missing |

*C. Test Summary*

We conducted multiple measurements on a real network, and the worst cases are depicted in Table III and Table IV, other runs were more accurate and aligned with starting event. The main causes of timestamp discrepancy are: 1) built-in jitters, which avoid alignment of several timeout events at the same time; 2) control-plane processing; 3) real device hardware processing. Different fast-start implementation by Cisco (which is not in compliance with the standard, see above) is the cause of missing LLDP messages in real network scenarios.

Validation discovered reasonable differences between our developed modules and referential implementation. The main goal of adding two new protocols to OMNeT++ was achieved.

## V. Conclusion

In this paper, we briefly described two most deployed Layer 2 discovery protocols – CDP and LLDP. We created simulation modules of these protocols within OMNeT++ discrete-event simulator as new software contributions. We tested and verified functionality and accuracy of our models in comparison with the real network running referential implementation.

It is valuable to support CDP and LLDP within (ANSA)INET not only for the sake of completeness of simulated network behavior but also for any future research efforts. Both protocols already employ TLVs, which allow very convenient way how to add new functionality. Hence, our implementation offers a great starting point for any proof-of-concept extending original protocols. For instance, Software Defined Network related use-cases and technologies offer an interesting playground for our framework.

More information about the ANSAINET project is available on the homepage [11]. All source codes including CDP and LLDP implementations could be downloaded from GitHub repository [12].

## Acknowledgment

This work was supported by the Brno University of Technology organization and by the research grant FIT-S-14-2299.

TABLE V. A LIST OF CDP AND LLDP TLVs

| CDP TLV | TLV's Description | LLDP TLV |
|---|---|---|
| Version | CDP protocol revision number. | |
| | Unique identifier of the device in the scope of local area network, which may be derived from Layer 2/3 address, chassis or port component number, etc. | Chassis Id |
| Time To Live | Information is stored in a neighbor table for a period specified by this TLV record. For CDP, recommended value is 3× longer than a periodic generation; for LLDP, it is 4× longer. | Time To Live |
| Checksum | Message content integration check computed similarly as IP header checksum. | |
| Address | TLV contains sender's address. Optionally, it may carry also reflected recipient's address | Management Address |
| Capabilities | Specifies device's role within a network such as a router, switch, bridge, etc. | System Capabilities |
| Port-Id | String representation of sender's interface port label including index. This TLV is handy for checking the improper cabling | Port Id |
| | The label is specifying additional information about the interface for administrative purposes. | Port Description |
| Full/Half Duplex | Duplexness of sender's interface. This information may be used to detect duplex mismatch between devices | |
| Native VLAN | TLV hosts configured native (untagged) VLAN on a trunk interface. This TLV may be used to detect native VLAN misconfiguration | |
| Device-Id | Device's hostname (e.g., router1.local.lab) | System Name |
| Location | Device's topology location (e.g., Omega Bld., Rack 1) | System Description |
| Platform | Device's hardware descriptor (e.g., Catalyst 3560) | |
| Software Version | Device's operating system information usually as multi-line string representation | |
| VTP Management Domain | VLAN management extension governing the borders of another Cisco's proprietary protocol called VLAN Trunking Protocol | |
| IP Network Prefix | On-demand routing extension of CDP suitable for hub-and-spoke topologies. This TLV carries a list of device's network segments and configured default gateway | |
| | The last TLV in the list marking the end of LLDP message. | EndOfLLDPDU |